# FLOW CONFIGURATION INFLUENCE ON DARCIAN AND FORCHHEIMER PERMEABILITIES DETERMINATION

Hussain NAJMI[1,*], Eddy EL TABACH[2], Khaled CHETEHOUNA[1], Nicolas GASCOIN[1], François FALEMPIN[3]

[1] INSA-CVL, Univ. Orléans, PRISME EA 4229, F-18022 Bourges, France
[2] Univ. Orléans, INSA-CVL, PRISME EA 4229, F-18000 Bourges, France
[3] MBDA-France, 8 rue le Brix, 18000 Bourges, France

[*] Corresponding author: hussain.najmi@insa-cvl.fr

**Abstract**

Within the framework of fuel cells, porous materials are used for filtration purpose. Determining physical properties like porosity and permeability are of utmost importance to predict and to manage filtration efficiency of the materials. The permeability of materials is often determined experimentally in laboratory with disc samples (outlet of the flow is achieved through the porous material) that are not exactly similar to the tubes which are used in realistic operating conditions (a main flow is found additionally to the one through the material). Thus, the effect of a second outlet on the Darcian's permeability characterization bench should be studied. In the present paper, we present a new test bench to determine experimentally the Darcy's and Forchheimer's permeabilities for a porous media by taking into account two outlets. Operating parameters (temperature, pressure and mass flow rate) are measured for three different configurations: *i*) secondary outlet (S.O) is 0% open, *ii*) S.O is 50% open and *iii*) S.O is 100% open. Then Darcy's and Forchheimer's permeability for all these three cases are compared and discussed in details. It has been found that the secondary outlet opening does not affect the Darcian's permeability but have substantial influence on the Forchheimer's one through the modifications of the range of study (lower mass flow rate penetrates the porous medium).

**Keywords:** *Filtration*; *Darcy's and Forchheimer's permeabilities*; *Porous stainless steel*; *Fuel cell application.*

## 1. Introduction

There are three main technologies for hydrogen separation which include: i) pressure swing adsorption (PSA), ii) cryogenic distillation and iii) selective permeation through a membrane. For fuel cell applications, $H_2$ separation by selective permeation through membrane has several advantages compared to the first two methods. Indeed, several literature works mention that this filtration technique is a simple operation which is characterised by a low cost and a less energy consumption as well as a high purity of hydrogen (Spillman et al. 1989, Lu et al. 2004, Phair et al. 2006, Chen et al. 2009 and Chi et al. 2010). This filtration process related to fuel cell can be found in other fields of application such as geology (Luquot et al. 2009, Abdulgader et al 2013), petrochemical (Tanaka et al. 2005, Shamsabadi et al. 2013), combustion (Franz et al. 2013, Burrerio et al. 2014) or aeronautics. For example, there have been large numbers of literature regarding transpiration cooling in aeronautics (Gascoin 2011, Wang et al 2011, Zhao et al. 2014). Performance criteria required for the membrane are high flux rate, stability at variable temperature, pressure and high selectivity even in the presence of other gases and contaminants (e.g. CO, $CO_2$, $CH_4$, $H_2O$ and $H_2S$ etc.). Temperature and pressure are the external parameters which can be controlled outside the porous media (i.e. membrane). But the selectivity depends upon the permeability of the material or of the porous medium (Thomasa et al. 2009 Baker et al. 2010).

It has been recognized in the past decade that the separation factor for gas pairs varies inversely with the permeability of the more permeable gas of the specific pair. Indeed, an analysis of the literature data for binary gas mixtures from the list of He, $H_2$, $O_2$, $N_2$, $CH_4$, and $CO_2$ reveals the above relationship for these mixtures (Robeson 1990, Chenar et al. 2006). Hence the determination of the permeability and studying the factor affecting it is of prime importance.

In previous works, a high pressure test bench with one inlet and one outlet configuration has been developed (Gascoin 2011, Fau et al. 2012). It has been tested on different type of metal and composite samples to determine the Darcian's and Forchheimer's permeabilities for inert and reactive fluids at high temperature and pressure. Similar experimental setup has been used by several authors to determine the permeability of disk and bar type specimens (Ilias et al. 1997, Gao et al. 2005, Papaioannou et al. 2015). Ilias et al. (1997) studied the permeability and selectivity of hydrogen through a palladium membrane at high temperature and low pressure. Gao et al. (2005) investigated the hydrogen permeability on porous stainless steel (PSS) disks coated with a mesoporous palladium impregnated zirconia intermediate layer. Papaioannou et al. (2015) evaluated carbon dioxide permeability for sintered supports with molten carbonate dual-phase membranes and its performances were found to be stable for more than 200 h of operation.

Despite of so many studies on the porous media, still we face a lack in evaluating the correct flow through porous media. Permeability of the material is mostly determined through experiments using disc sample in laboratory. The outflow through the porous media is measured to determine the permeability of the material. But it is not exactly similar to a realistic condition when fluid flows through a tube. In such case, a perpendicular flow to the main one occurs through the porous medium (cf. Figure 1). Thus, it makes necessary to study the effect of second outlet on the determination of Darcian's and Forchheimer's permeabilities (despite the fact that permeability is often considered as a characteristic of the material itself Allaby et al 1999).

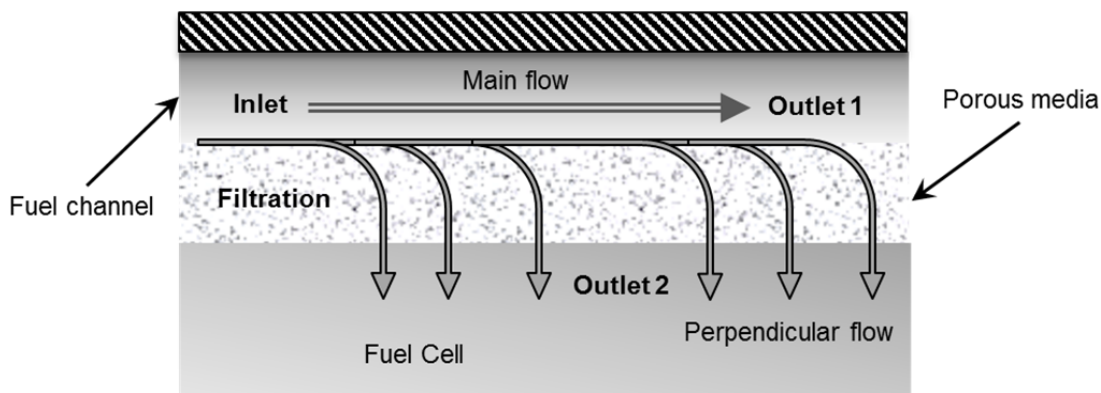

Figure 1: Overview of realistic flows within channel surrounded by porous medium.

The aim of this paper is to improve previously used experimental setup in order to determine accurate Darcian's and Forchheimer's flow behaviour inside a porous media in a realistic configuration. The effect of a second outlet on the Darcian's and Forchheimer's permeabilities is investigated. The next section is devoted to the description of the adapted test bench where different opening of second outlet is studied. The Darcian's and Forchheimer's permeabilities of porous stainless steel materials of class 3 (SS3) and class 20

(SS20) are measured and the obtained results are presented and discussed in the third section of this paper.

## 2. Materials & Methods

*2.1. Permeation experimental and test methodology*

The bench used here is composed of a cylindrical permeation cell connected to gas injection system (cf Figure 2). Porous material samples are placed in this permeation cell which is composed of two main parts (High Pressure Chamber: HPC for the inlet and Low Pressure Chamber: LPC for the outlet) in order to maintain the porous media in the fluid flow and to avoid leakage. Despite its small size (external diameter of 40 mm), one inlet and two outlet (outlet-1 and outlet-2) enable to measure the pressure and mass flow rate on each side of the porous sample. The outlet-2 is considered as a Primary Outlet (P.O) whereas the outlet-1 is considered as secondary outlet (S.O). The injection system is capable of handling liquid as well as gaseous fuels with a high pressure pump (80 bars, 0.5 $g.s^{-1}$). Nitrogen gas is used as fluid. Four sensors to monitor transient variations of mass flow rate, pressure and temperature are connected to a data acquisition system (about 1 Hz, 16 bits, 48 channels). The mass flowmeter with a range of 0 to 3 g/s is placed at outlet-2 in order to measure mass flow rate of fluid through porous media. Pressures are measured at inlet and outlet-2 to get the pressure in HP chamber ($P_{inlet}$) and pressure difference across the sample. Pressure range of a sensor at inlet side is 0 to 60 bar whereas ΔP can measured in a range of 0 to 635 mbar. Flow from the secondary outlet is controlled by manually operated valve where the three opening percentage values (0%, 50% and 100%) are selected. Figures 2 and 3 illustrate respectively a photograph and a schematic overview of the developed permeation experimental setup.

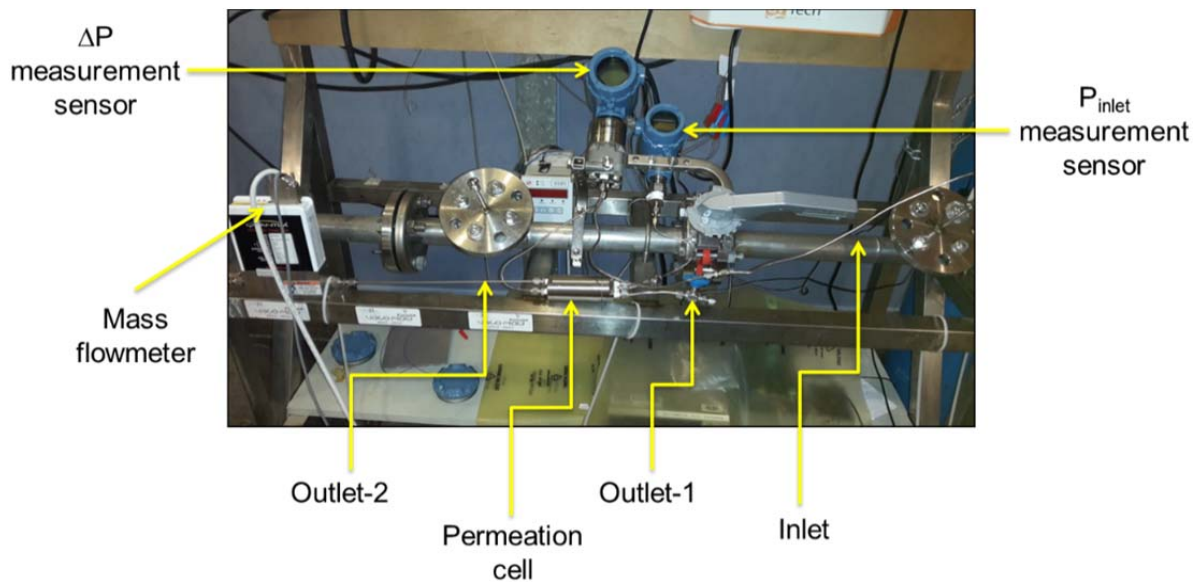

Figure 2: Photograph of the permeation test bench.

Permeability measurements are performed on two different classes of stainless steel that is SS3 and SS20 (cf. Figure 4). Specifications of both porous materials used are given in Table 1 (Gascoin 2011). Except one consideration that the outlet of permeation cell should be at atmospheric pressure during the experiments, the norm ISO 4022 for the determination of fluid permeability is followed. Indeed, due to the effect of secondary outlet it makes necessary to place the mass flowmeter at primary outlet (result in back pressure) so that the flow through porous media can be accurately measured and permeabilities of the porous media can be determined. All tests are performed at room temperature and repeated at least three times.

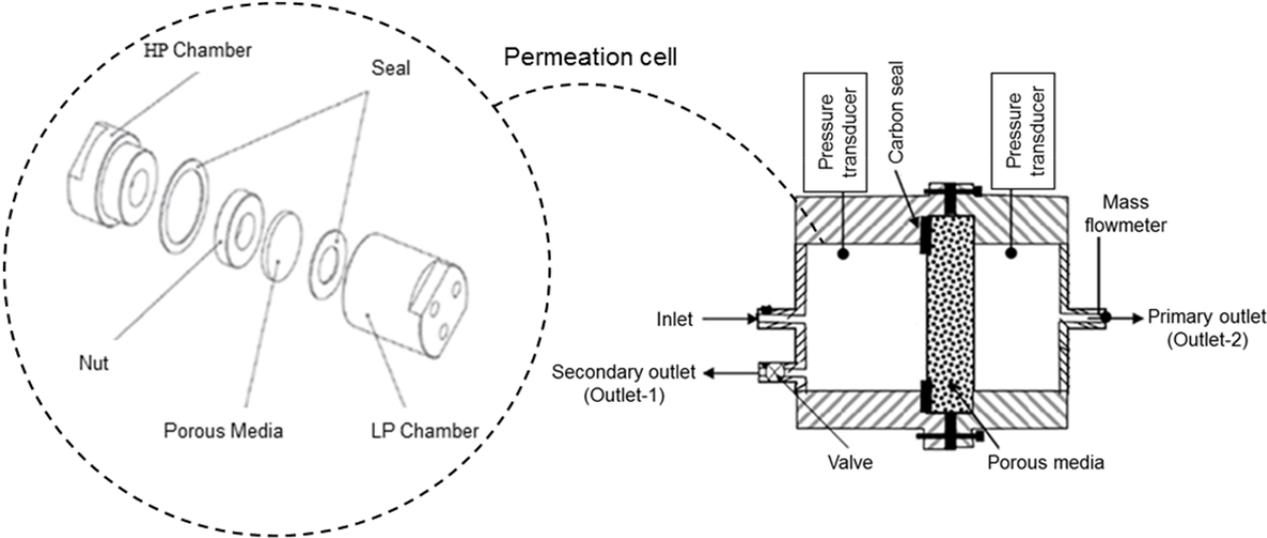

Figure 3: Schematic overview of the permeation experimental setup.

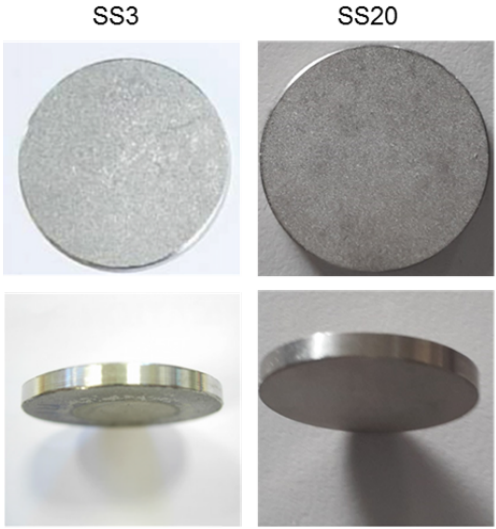

Figure 4: Stainless steel porous medium: Fe (62 wt.%), Cr (18 wt.%), Ni (11 wt.%) C (7 wt.%), O (1 wt.%) and Si (1 wt.%).

| Type | SS3 | SS20 |
| --- | --- | --- |
| Diameter | 30mm | 30mm |
| Filtration Diameter | 16mm | 16mm |
| Thickness | 3mm | 3mm |

| Overall Porosity | 30% | 36% |
| Composition | Fe (62 wt.%), Cr (18 wt.%), Ni (11 wt.%) | C (7 wt.%), O (1 wt.%) and Si (1 wt.%) |

Table 1. Specifications of the two porous stainless steel materials.

*2.2. Determination procedure of Darcian's and Forchheimer's permeabilities*

Accurate description of fluid flow behavior in the porous media is essential and is described in several ways in various literature studies. Usually, Darcy's law depicts fluid flow behavior in porous media. According to Darcy's law, the pressure gradient is linearly proportional to the fluid velocity in the porous material. This linear relationship is valid only for very small flow velocities or low pressure gradients. As the Reynolds number (Re) approaches to a critical value (i.e. turbulent regime), the relationship becomes nonlinear. The one-dimensional Darcy equation can be written as:

$$\frac{dP}{dx} = -\frac{\mu V}{K} \quad (1)$$

where $P$ is the pressure, $x$ is the direction of fluid flow, $\mu$ is the fluid viscosity, $V$ is the superficial velocity, and $K$ is the permeability. The transient behavior in porous media is described by Navier–Stokes equations whereas Brinkman's equation below is used to describe the macroscopic fluid flow in large range flow regimes:

$$\frac{\Delta P}{L} = \mu \frac{V}{K_D} + \rho \frac{V^2}{K_F} \quad (2)$$

where $\Delta P = P_{inlet} - P_{outlet}$ is the pressure drop through the porous medium, $L$ is the external mean sample thickness, $\rho$ is the inlet density (with respect to inlet pressure), $K_D$ and $K_F$ are the Darcy's and Forchheimer's terms. The right term of Eq. (2) is composed of two parts, one related to the Darcy's law for low velocity regime filtration (Darcian flows) and the quadratic one is related to the Stokes's law (non-Darcian flows) which takes into account the inertial effects related with flow resistance. The Stokes equation is also called as Forchheimer's equation (Tully et al., 2005 and Murthy and Singh, 2000). Numerous other formulations of the Brinkman's equation can be found (Kim and Park, 1999, Choi et al., 1998, Martin and Boyd, 2008 and Valdes-Parada et al., 2007). Large range of flow regimes through porous media can also be described by using power laws (Rathish and Shalini, 2003) and cubic laws (Aulisa et al., 2009, Nguyen et al., 2007 and Pazos et al., 2009).

In order to compute Darcian's and Forchheimer's permeabilities from the measured parameters on the experimental bench, a modified form of the above equation is derived:

$$\frac{\Delta p}{L \mu V} = \frac{1}{K_D} + \frac{1}{K_F} \frac{\rho V}{\mu} \quad (3)$$

Thus, the left term $\frac{\Delta p}{L\mu V}$ is plotted as a function of $\frac{\rho V}{\mu}$. The origin of the plot is linked to the Darcian's term while the angle of climb or slope of the plot is related to the Forchheimer's term. Due to the implication of density and dynamic viscosity in the pressure drop formula of Eq. (3), their estimation is of great importance. The density is computed on the basis of measured pressure and temperature thanks to the modified perfect gases law with the compressibility factor Z which depends notably on the critical coordinates and on the Pitzer acentric factor (Gascoin 2011). The dynamic viscosity is computed by the method proposed by Chung (Poling et al. 2001).

## 3. Results and discussion

As mentioned in previous sections, the Darcian's and Forchheimer's permeabilities of two kind of stainless steel porous materials are determined and the influence of different openings of secondary outlet is investigated. Before proceeding, some preliminary experiments are performed on SS3 in order to verify that the placement of mass flowmeter at primary outlet instead of inlet, as per ISO 4022 norm, does not affect the permeabilities estimation. Results of Darcian's permeability in the two cases (i.e. mass flowmeter at primary outlet and mass flowmeter at inlet) show clearly that the relative gap do not exceed 8% whereas those of Forchheimer's permeability is inferior to 13% (these values are in the range of previously estimated accuracy, Gascoin 2011). These relative gap values confirm that the placement of mass flowmeter at primary outlet do not affect the determination of permeabilities. When Brinkman's equation is considered (turbulent flow), the Darcy's term for carbon cooled porous structure is found to range from $10^{-13}$ to $10^{-12}$ m$^2$ and the Forchheimer's coefficient ranges from $10^{-8}$ to $10^{-7}$ m (Langener et al. 2008).

The Darcian's and Forchheimer's permeabilities are determined with the plots of the term $\frac{\Delta p}{L\mu V}$ a function of $\frac{\rho V}{\mu}$ for the both studied stainless steel porous media which are illustrated in Figure 5 for different opening of secondary outlet (0%, 50% and 100%).

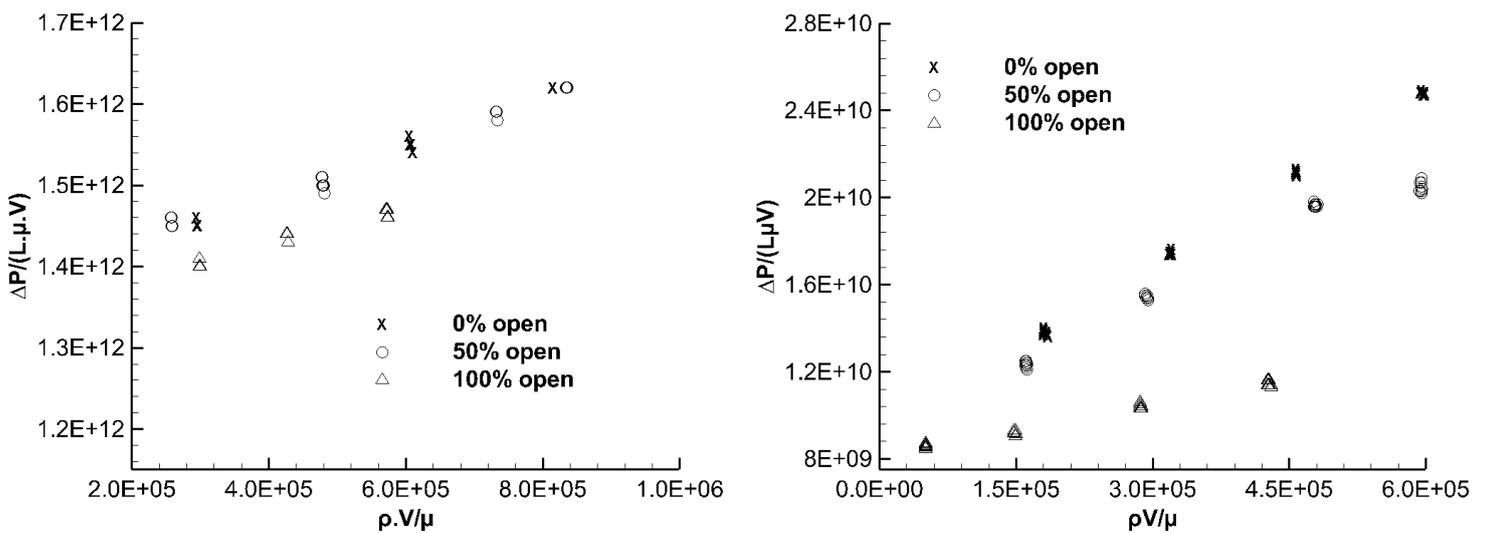

Figure 5. Determination of Darcian's and Forchheimer's permeabilities for different opening of secondary outlet. Left: SS3 and Right: SS20.

Using the results plotted in above Figure 5, the intercept of the different lines on Y axis will give the $K_D$ coefficients and their slopes lead to $K_F$ coefficients. The obtained values of Darcian's and Forchheimer's permeabilities are given in Table 2 for different openings of secondary outlet. It can be clearly seen from this Table that in case of both the porous materials (SS3 and SS20) the determination of $K_D$ is not affected by the opening of the secondary outlet. For SS3 porous material, a discrepancies of 1.21% and 0.1% are respectively found for 50% and 100% open cases. Similarly in case of SS20 porous media, relative gaps of 5.4% and 10.8% are respectively obtained for 50% and 100% opening of secondary outlet. But when it comes to determination of $K_F$, the opening of secondary outlet seems to have an effect. Indeed, relative changes of 11% and 71% are found in case of SS3 whereas values of 34% and 237% are obtained for SS20 for 50% and 100% open cases respectively. $K_F$ is varying a lot for the 100% cases, it is because of the fact that the flow through the porous media is very low that results in decrease of pressure drop which is linearly proportional to square of velocity at higher flow rates (Non Darcian Flow). Since $K_F$ coefficient is associated to the turbulent flow inside the porous media (Firoozabadi et al. 1979, Phair et al 2006, Lemos 2012), hence the velocity of fluid in the material is too low and the flow is not or less turbulent and it makes its determination less accurate. That's is why the value for the $K_F$ is quite far from the "real" one.

| Type of material | Permeability Term | Secondary outlet opening percentage | | |
|---|---|---|---|---|
| | | 0% Open | 50% Open | 100% Open |
| SS3 | $K_D$ (m$^2$) | 7.38×10$^{-13}$ | 7.29×10$^{-13}$ | 7.37×10$^{-13}$ |
| | $K_F$ (m) | 3.08×10$^{-06}$ | 3.43×10$^{-06}$ | 5.28×10$^{-06}$ |
| SS20 | $K_D$ (m$^2$) | 1.11×10$^{-10}$ | 1.05×10$^{-10}$ | 1.23×10$^{-10}$ |
| | $K_F$ (m) | 3.79×10$^{-05}$ | 5.11×10$^{-05}$ | 1.28×10$^{-04}$ |

Table 2. Darcian's and Forchheimer's permeability values at three different secondary outlet openings for SS3 and SS20.

In order to give a deeper explanation to the above statement and study the influence of Darcian's and Forchheimer's contribution on pressure losses, Figure 6 presents the evolutions of these contribution as well as their ratio as functions of pore Reynolds number. The pore Reynolds number has been computed for metallic samples by using the relation $R_e = \rho V d_p / \mu$ where $d_p = 4\varepsilon / (a_g (1-\varepsilon))$ is the pore diameter, $\varepsilon$ is the porosity of material, $a_g = 6/d_g$ is the grain area and $d_g$ is grain diameter. Let us notice that the grain diameter of the SS3 sample is measured in a previous work (Gascoin 2011) $d_g = 14.1\,\mu m$ whereas the one of the SS20 porous media $d_g = 55.2\,\mu m$ is computed from a correlation between class and grain diameter deduced from the literature data. It is observed from the set of plotted data that in case of SS3 porous media Darcian's term is always higher than the Forchheimer's term for all the three different opening of secondary outlet (0%, 50% and 100%) showing that the flow is always in a laminar regime. There is no influence of secondary outlet opening on Darcian's term and its contribution remains constant for all the three cases but Forchheimer's contribution decreases as the secondary outlet opening increases. The similar trend is observed in case of SSS20 except that in case of 0% and 50% opening of secondary outlet where the Forchheimer's contribution overcomes the Darcian's contribution as the pore Reynolds number exceeds a critical value of $R_{e_c} = 6$. It is also observed that the decrease in Forchheimer's term for SS3 porous material is less evident compare to the Forchheimer's decrease in case of SS20.

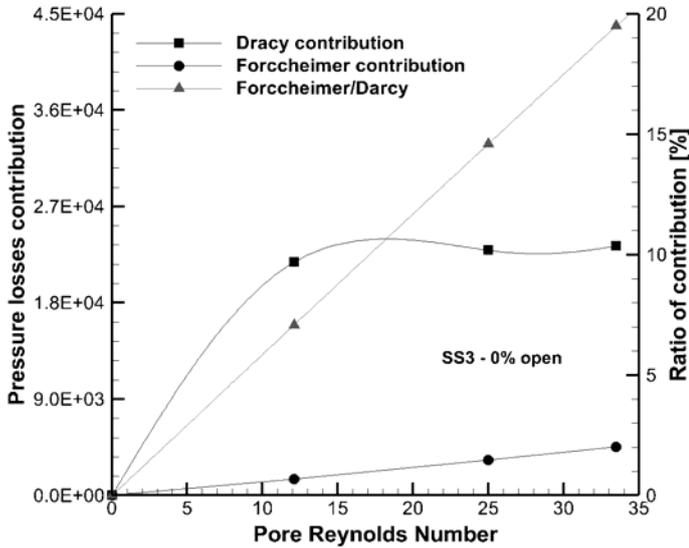
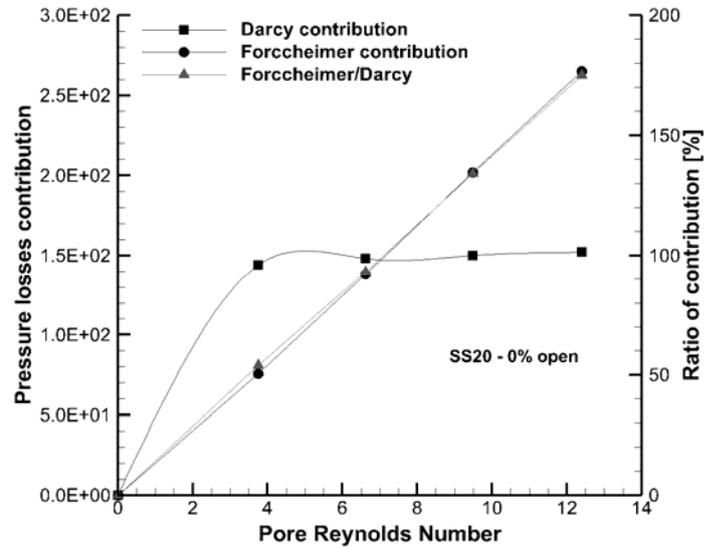
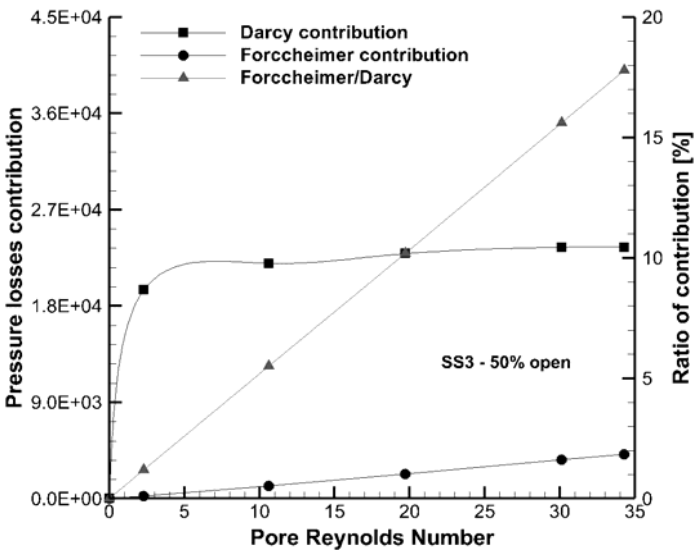
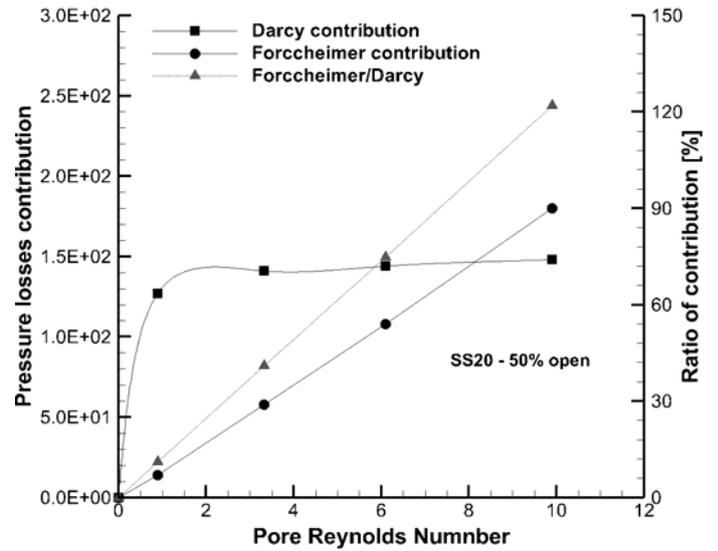

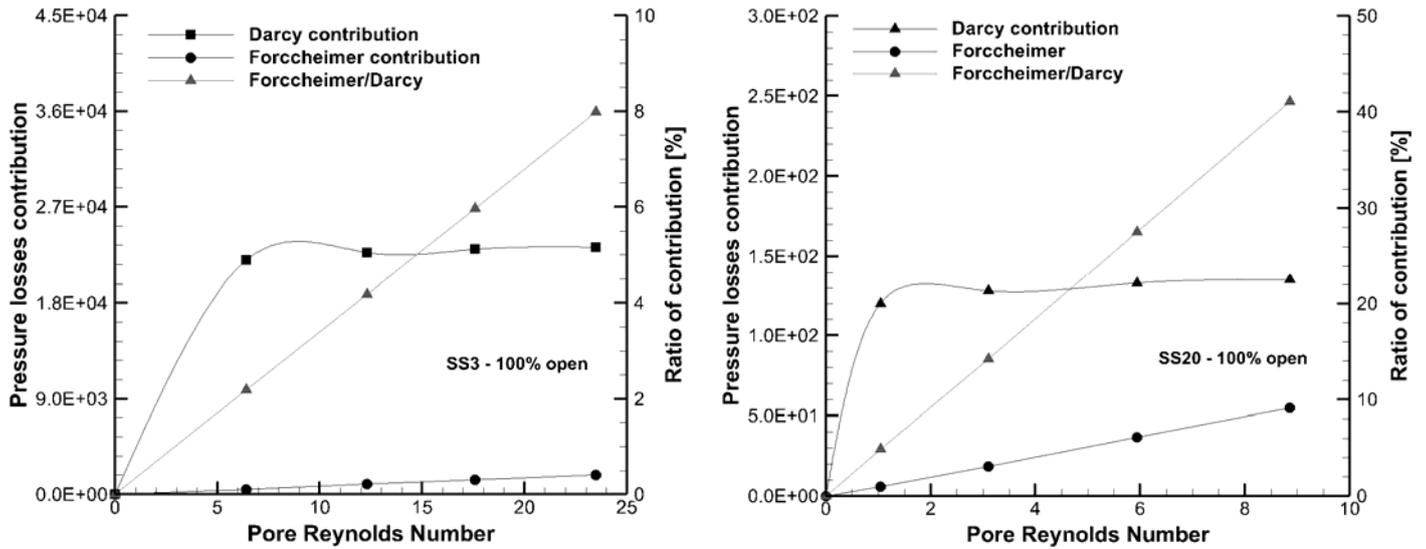

Figure 6. Permeation contributions plotted as a function of the estimated pore Reynolds number for class 3 and class 20 for different openings of secondary outlet.

## 4. Conclusion

In fuel cell applications, porous materials is used for the filtration need. Determining physical properties like porosity and permeability are of utmost importance to predict and manage filtration efficiency of the materials. An experimental test bench was design to determine both Darcian's and Forchheimer's permeabilities. Nevertheless, it was found later that using the experimental values of these permeabilities on complete integrated fuel reforming systems led to discrepancies of several orders of magnitude. One of the hypothesis that was pose is that the geometrical configuration may impact the results. In other words, having one flow inlet and one flow outlet or one inlet and two outlets (as it was discussed in this paper) may lead to different apparent permeabilities. Effect of secondary outlet on permeability determination has thus been investigated. It was found that there is no much change is observed in $K_D$ value when the mass flowmeter is placed at outlet-2 instead of the inlet. When Darcian's permeability is determined for SS3 and SS20 with different openings for secondary outlet, negligible changes are observed, hence they show that this term is not affected by change in flow configuration in case of stainless steel material. But secondary outlet opening have significant effect on the determination of $K_F$ value for both kinds of porous material. Indeed, its criterion of determination is highly governed by turbulent flow regime. This flow regime is found in just two cases out of six cases and that is only when the pore Reynolds number exceed its critical value.

In future works, Darcian's and Forchheimer's permeabilities will be studied in a tube configuration using different composites. Mathematical background will be developed in order to correlate these permeabilities to the selectivity of the porous materials. After that selectivity of several binary mixtures will be investigated on different porous media using an adapted experimental setup.